\documentclass[epj, twocolumn]{webofc}
\usepackage[varg]{txfonts}   
\woctitle{The Innermost Regions of Relativistic Jets and Their Magnetic Fields}
\begin{document}
\title{Gamma-ray emission from early-type stars interacting with AGN jets}
%
%

\author{Anabella T. Araudo\inst{1}\fnsep\thanks{\email{a.araudo@crya.unam.mx}} 
\and
Valent{\'\i} Bosch-Ramon\inst{2}\fnsep \and Gustavo E. Romero\inst{3,4}\fnsep}

\institute{Centro de Radioastronom\'{\i}a y Astrof\'{\i}sica, 
Universidad Nacional Aut\'onoma de M\'exico, A.P. 3-72 (Xangari), 
58089 Morelia, Michoac\'an, M\'exico.
\and
Universitat de Barcelona. Departament d'Astronomia i Meteorologia
Marti i Franques 1, 7th floor ES 08028 Barcelona, Spain.           
\and
Instituto Argentino de
Radioastronom\'{\i}a, C.C.5, (1894) Villa Elisa, Buenos Aires, Argentina.
\and
Facultad de Ciencias Astron\'omicas y Geof\'{\i}sicas,
Universidad Nacional de La Plata, Paseo del Bosque, 1900 La Plata, Argentina.}

\abstract{
We study the interaction of early-type stars with
the jets of active galactic nuclei. A bow-shock  will form as a consequence 
of the interaction of the jet with the winds of stars and
particles can be accelerated up to relativistic
energies in these shocks. We compute the non-thermal radiation produced by 
relativistic electrons from radio to gamma-rays. This 
radiation may be significant, and its detection might yield
information on the properties of the stellar population in the galaxy
nucleus, as well as on the relativistic jet. This emission is expected 
to be relevant for  nearby non-blazar sources. 
}
\maketitle
\section{Introduction}
\label{intro}

Active galactic nuclei (AGNs) consist of a supermassive
black hole (SMBH) surrounded by an accretion disc in the center of a galaxy. 
Sometimes these objects 
present radio emitting jets originated close to the SMBH and ejected 
perpendicular to the accretion disc. 
Radio-loud AGNs present thermal and non-thermal continuum  emission 
in the whole electromagnetic spectrum, from radio to gamma-rays. 

In the nuclear region of AGNs there is matter in the form of
diffuse gas, clouds, and stars, making jet medium interactions
likely \citep{Araudo_10}. We study the interaction of 
massive stars with the AGN jets \citep{Komissarov_04, Maxim_10}. 
We adopt the main idea of
\cite{Bednarek}, i.e. the interaction of massive stars with
AGN jets, although our scenario consists of a population of massive
stars surrounding the jets, and considers jet-star interactions at
different heights ($z$) of the jet. We analyze the dependence with $z$
of the properties of the interaction region (i.e. the shocks in the
jet and the stellar wind), and also the subsequent non-thermal
processes generated at these shocks. We consider the injection of 
relativistic electrons, the evolution of this population
of particles by synchrotron and inverse Compton (IC) radiation,
as well as escape losses.

In the scenario considered here, the emitters are  the flow
downstream of the bow shocks located around the stars. This flow moves
together with the stars at a non-relativistic speed, and thus the
emission will not be relativistically boosted. For this reason the
radiation from jet-star interactions will be mostly important in
misaligned AGNs, where the emission produced by other mechanisms in the
jet is not amplified by Doppler boosting. 
In the GeV domain, the {\it Fermi} satelllite has already detected at least 11 
misaligned radio-loud AGNs, a population that is expected to grow in the near
future. Because of this, theoretical models that can predict the level
and spectrum of the gamma-ray emission from these sources are desirable
to contribute to the understanding of future detections.

\section{Stellar populations in AGNs}

The number of stars formed per unit of mass ($m$), time ($t$) and volume
($V \propto r^3$) is  $\psi(m,r,t) = \psi_0(m,r)
\exp(-t/T)$,  where $t$ and $T$ are the age of the stellar system  and the
duration of the formation process, respectively.  We consider
that stellar formation processes take place continuously ($t \ll T$) 
in the nuclear region of the galaxy, and 
the stars are uniformelly distributed around the SMBH.
We assume that 
\begin{equation}
\psi = K \left(\frac{m}{M_{\odot}}\right)^{-2.3}
\left(\frac{r}{\rm pc}\right)^{-2}, 
\end{equation}
where $0.1 \leq m/M_{\odot} \leq 120$. (In \citep{Araudo_13}, 
$\psi \propto r^{-1}$ is also considered.) The star formation rate is
$\dot M_{\star} = \int\int \psi \,m \,{\rm d}m \,{\rm d}V$, i.e.:
\begin{equation}
\label{M_dot_M}
\dot M_{\star} =  K 
\int_{1 \rm pc}^{1 \rm kpc} \left(\frac{r}{\rm pc}\right)^{-2} 4\pi r^2 {\rm d}r 
\int_{0.1 M_{\odot}}^{120 M_{\odot}} \left(\frac{m}{M_{\odot}}\right)^{-1.3} {\rm d}m.
\end{equation}
%
We integrate Eq.~\ref{M_dot_M} from 1~pc to 1~kpc, which is the maximun 
value where the following empirical relation  is valid \citep{Satypal_05}: 
\begin{equation}
\label{M_dot}
\frac{\dot M_{\star}}{\rm M_{\odot}\,yr^{-1}} \sim
47.86 \left(\frac{\dot M_{\rm bh}}{\rm M_{\odot}\,yr^{-1}}\right)^{0.89} \sim
714\, \eta_{\rm acc}^{0.89}.
\end{equation}
In the previous equation we have used that 
the SMBH accretion rate, $\dot M_{\rm bh}$, is related with the SMBH mass,
$M_{\rm bh}$, as $0.1 \dot M_{\rm bh} c^2 = \eta_{\rm acc} L_{\rm Edd}$, where  
$L_{\rm Edd} = 1.2\times10^{47} (M_{\rm bh}/10^9\,M_{\odot})$~erg~s$^{-1}$
is the Eddington luminosity. 
We fix $M_{\rm bh} = 10^9$~M$_{\odot}$,  and $\eta_{\rm acc} = 0.01$,
0.1, and 1. (In \citep{Araudo_13} different values of $M_{\rm bh}$ are 
considered.)
Then, from Eqs.~(\ref{M_dot_M}) and (\ref{M_dot}), $K$ results 
$\sim 0.01 \,\eta_{\rm acc}^{0.89}$.

\begin{figure}
\centering
\includegraphics[angle=270, width=0.5\textwidth]{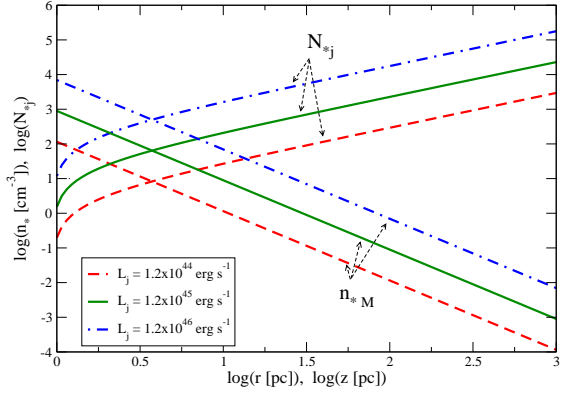}
\caption{Density of massive stars ($n_{\star \rm M}$) and number  of massive 
stars inside the jet ($N_{\star \rm j}$) for the different values of $z$. 
Cases for different values of $\eta_{\rm acc}$ are plotted.}
\label{population}       
\end{figure}

Once a stellar population is injected in the host galaxy, stars of a given 
mass are accumulated in the galaxy  and, at a time
$t < t_{\rm life}$, where $t_{\rm life} = a (m/M_{\odot})^{-b}$~Gyr 
is the stellar lifetime, the density of stars (per unit of mass) is 
$n_{\star m} \approx \psi\,t$. For $t>t_{\rm life}$, stars die and the mass
distribution follows a law $n_{\rm \star m}\propto m^{-(2.3+b)}$. 
In the case of massive stars, $(a,b) = (1,1.7)$ and $(0.1,0.7)$
for $m < 15$~M$_{\odot}$  and  $60 > m/M_{\odot} > 15$, respectively 
\citep{Ekstrom_12}. For $m > 60$~M$_{\odot}$, $t_{\rm life} \sim 0.004$~Gyr. 
Then, at
$t \gtrsim t_{\rm life}(8 M_{\odot}) \sim 0.03$~Gyr, the rate of stellar formation
is equal to the rate of stellar death and the system reaches the steady 
state for $m>8\,M_{\odot}$. In such a case, the number density of
massive stars -$n_{\star \rm M}$- keeps the spatial  dependence
of the stellar injection rate, $\psi \propto r^{-2}$, resulting 
\begin{equation}
\label{n_masivas}
\frac{n_{\star \rm M}}{\rm pc^3}  = 
\int_{8\,M_{\odot}}^{120\,M_{\odot}}n_{\rm \star m}\,{\rm d}m \sim 
6.9\times10^3\,\eta_{\rm acc}^{0.89}\left(\frac{r}{\rm pc}\right)^{-2}
\end{equation}
as is shown in Fig.~\ref{population}.

The number of massive stars contained in the jet volume is  
$N_{\rm \star j}(z) = \int_{1\,{\rm pc}}^{z} n_{\star\rm M}(z'){\rm d}V_{\rm j}$, 
where d$V_{\rm j} = \pi R_{\rm j}^2{\rm d}z'$ 
($z$ is the $r$-coordinate along the jet and the jet radius $R_{\rm j}$ is
$\sim 0.1 z$). This yields: 
\begin{equation}
\label{N_stars}
N_{\rm \star j} \sim 8.6\times10^4\,\eta_{\rm acc}^{0.89}
\left[\left(\frac{z}{\rm pc}\right) -1\right].
\end{equation}
Note that at $z\ge z_1 \sim \eta_{\rm acc}^{-0.89}$~pc
there is at least one massive star inside the jet at every time (see 
Fig.~\ref{population}).

\section{Jet-star interaction}
\label{sec-1}

We consider  massive stars with mass loss rate and 
terminal wind velocity $\dot M_{\rm w} = 10^{-6}$~M$_{\odot}$~yr$^{-1}$ and   
$v_{\infty} = 2000$~km~s$^{-1}$, respectively. 
When the jet interacts with stars a double bow shock is formed around them.
The location of the stagnation point is at a distance $R_{\rm sp}$
from the stellar surface, where the wind and jet ram pressures are equal.
From $\rho_{\rm w}\,v_{\infty}^2 = \rho_{\rm j}\,c^2\,\Gamma$, where
$\rho_{\rm w} \sim \dot M_{\star}/(4\pi R_{\rm sp}^2 v_{\infty})$ is 
the wind density, we obtain
\begin{equation}
\frac{R_{\rm sp}}{R_{\rm j}} \sim  
10^{-3}\left(\frac{\dot M_{\rm w}}{10^{-6}\,{\rm M_{\odot}/yr}}\right)^{1/2}
\left(\frac{v_{\infty}}{2000\,{\rm km/s}}\right)^{1/2}
\left(\frac{L_{\rm j}}{10^{44}\,{\rm erg/s}}\right)^{-1/2},
\end{equation}
resulting $R_{\rm sp} \propto z$.
We assume that the jet has a Lorentz factor $\Gamma = 10$ and
a velocity  $\sim c$. 
The jet kinetic luminosity is determined as $L_{\rm j} = \eta_{\rm j} L_{\rm Edd}$,
giving $L_{\rm j} = 1.2\times 10^{44}$,  $1.2\times 10^{45}$, and 
$1.2\times 10^{46}$~erg~s$^{-1}$,  for $\eta_{\rm j} = 0.001, 0.01$, and $0.1$, 
respectively. The jet density is obtained as 
$\rho_{\rm j} = L_{\rm j}/[(\Gamma -1) c^3 \pi R_{\rm j}^2]$.

\section{Particle acceleration}

Electrons are accelerated up to relativistic energies in both
the jet  and wind  shocks, and injected in the downstream regions following 
a distribution $Q_{e} \propto E_{e}^{-2.2}$. Under
the assumption of  a one-zone model for the accelerator/emitter, we solve 
\begin{equation}
\label{evolution} 
\frac{N_{e}}{t_{\rm esc}}-\frac{{\rm d}}{{\rm d} E_{e}}(\dot E_{e} N_{e}) =Q_{e}\,
\end{equation}
to derive the energy distribution of relativistic electrons $N_{e}$, 
where $t_{\rm esc} = \min\{t_{\rm adv}, t_{\rm diff}\}$.
The diffusion timescale is
$t_{\rm diff} \sim D_{\rm j,w}^2 q B_{\rm jbs,wbs}/(E_{e}\,c)$ in
the Bohm regime, where $B_{\rm jbs}$ and $B_{\rm wbs}$ are the magnetic fields in
the jet and the stellar wind bow-shock regions, respectively, and $q$ is the 
electron charge. The advection escape times in the
downstream regions of the jet and the wind bow shocks are
$t_{\rm adv,j} \sim 3\,R_{\rm sp}/c$ and $t_{\rm adv,w} \sim
4\,R_{\rm sp}/v_{\infty}$,  respectively.
Besides escape losses, electrons suffer  synchrotron and stellar photon 
IC upscattering losses, $\dot E_{e}$. 
For the later we have considered 
stellar target photons with an energy and luminosity $\sim 30$~eV and
$L_{\star} = 3\times10^{38}$~erg~s$^{-1}$, respectively. For synchrotron and
diffussion we have to estimate $B_{\rm jbs}$ and $B_{\rm wbs}$.

\begin{figure}
\centering
\includegraphics[angle=270, width=0.5\textwidth]{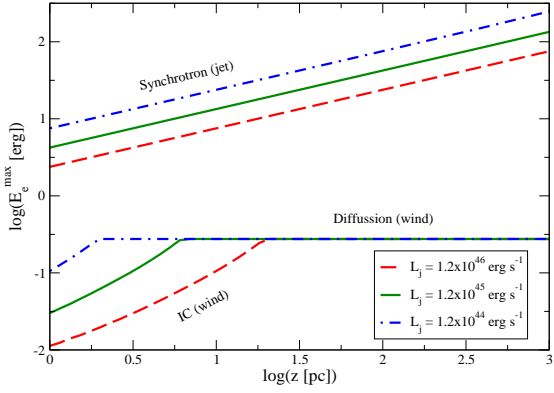}
\caption{Maximum energies of electrons accelerated in the jet (top)
and wind (bottom) bow shocks at different $z$.}
\label{Emax}       
\end{figure}

Assuming that the magnetic energy density in the jet is a fraction $0.3$ 
of its kinetic energy density \citep{Komissarov_07},   
and that in the shocked region the magnetic field is amplified by a factor
of 4, we obtain
\begin{equation}
B_{\rm jbs}  \sim 
\left(\frac{L_{\rm j}}{10^{44}\,{\rm erg\,s^{-1}}}\right)^{1/2}
\left(\frac{z}{\rm pc}\right)^{-1}\,\,{\rm G}.
\end{equation}
With this value of $B_{\rm jbs}$, the maximum energy achieved by electrons 
in the jet bow shock is determined by synchrotron losses, giving 
$E_e^{\rm max} \sim 
1.2\times10^2 (z/{\rm pc})^{1/2} (L_{\rm j}/10^{44}\,{\rm erg\,s^{-1}})^{-1/4}$~TeV.
For the wind we assume the parametrization of the 
magnetic field given in  \citep{Usov} , with a value
in the stellar surface of about 10~G. The maximum energies for electrons 
accelerated in the wind bow shock are determined by 
IC scattering and diffussion (see Fig.~\ref{Emax}).

Taking into account the escape,  synchrotron, and IC losses described
above, we solve Eq.~(\ref{evolution}) obtaining the energy
distribution $N_e$ of relativistic electrons in the jet and in the wind.  
In the former,  synchrotron and IC cooling dominates a
significant part of the electron energy distribution up to a certain
height, at which advection losses become dominant. (However, $E_e^{\rm max}$
in the jet is always constrained by synchrotron cooling.)
This is due to the different $z$-dependence of these
timescales, $z^2$ for synchrotron and IC, and $z$ for advection. 
In the later,  synchrotron and IC cooling dominates a
significant part of the spectrum, but at large values of $z$ 
diffusion losses become dominant all the way up to $E_e^{\rm max}$. 
In both cases, Thomson IC and synchrotron dominance 
appear as a steepening in $N_e$ from $\propto E_e^{-2.2}$ to 
$\propto E_e^{-3.2}$. 

\section{Non-thermal emission}

Once $N_e$  in the jet and wind shocked regions is computed,
we calculate the spetral energy distribution (SED) of the non-thermal
radiation, synchrotron and IC scattering (in Thomson and Klein-Nishina regimes) 
in the jet and the wind shocked regions,  using the standard fomulae. 
The energy budget for the emission
produced in the bow shock regions are 
$\sim \eta_{\rm nt} (R_{\rm sp}/R_{\rm j})^2L_{\rm j}$ 
and $\eta_{\rm nt} L_{\rm w}/4$,  where $L_{\rm w} = \dot M_{\rm w} v_{\infty}^2/2$
and $\eta_{\rm nt}$ is fixed in $0.1$. 
An important characteristic of the jet/star scenario  is
that the  emitter is fixed to the star, and being the star moving at a
non-relativistic velocity, the emission produced in the bow shock
regions is not amplified by Doppler boosting.

\begin{figure*}
\centering
\includegraphics[angle=270, width=0.49\textwidth,clip]{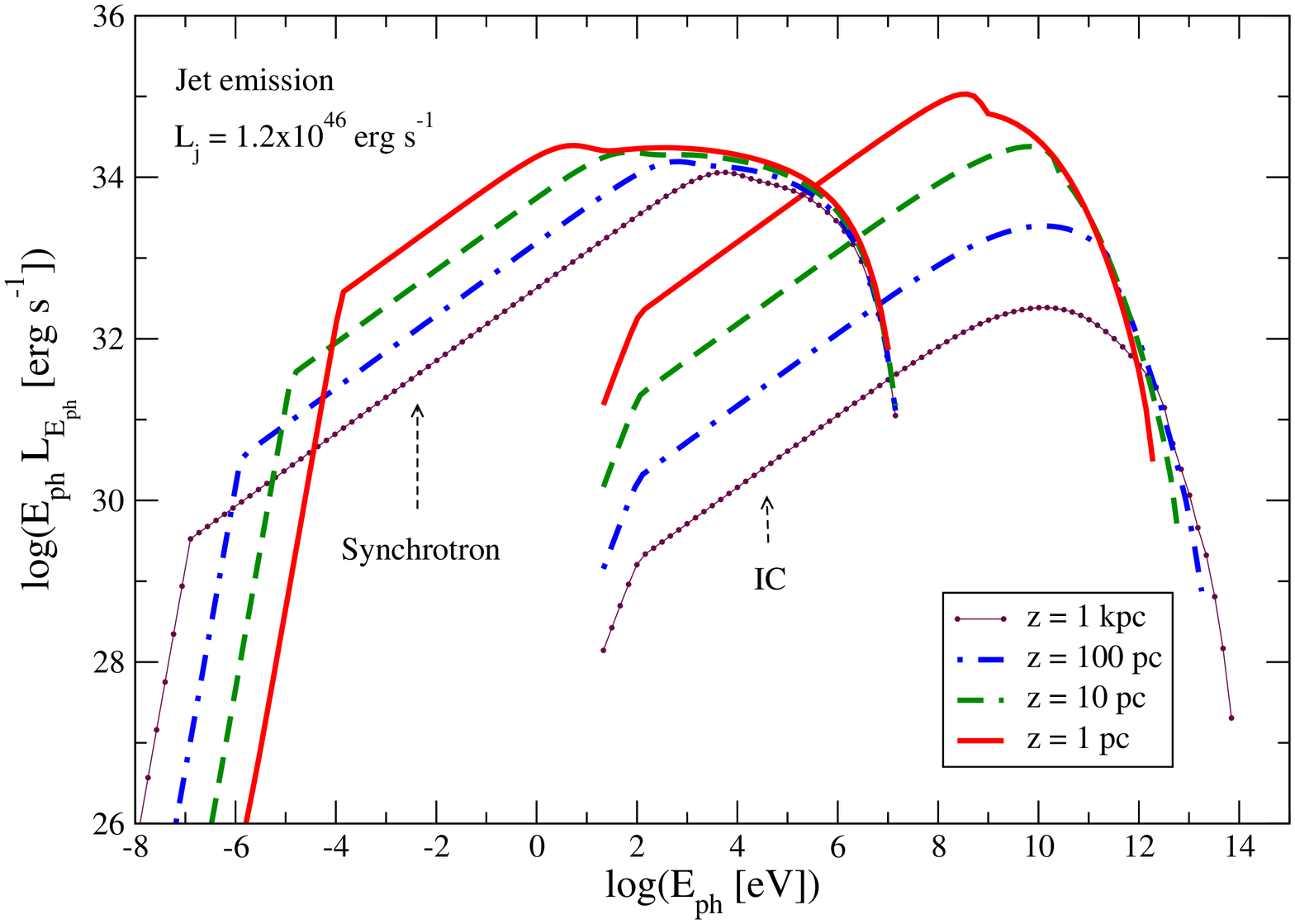}
\includegraphics[angle=270, width=0.49\textwidth,clip]{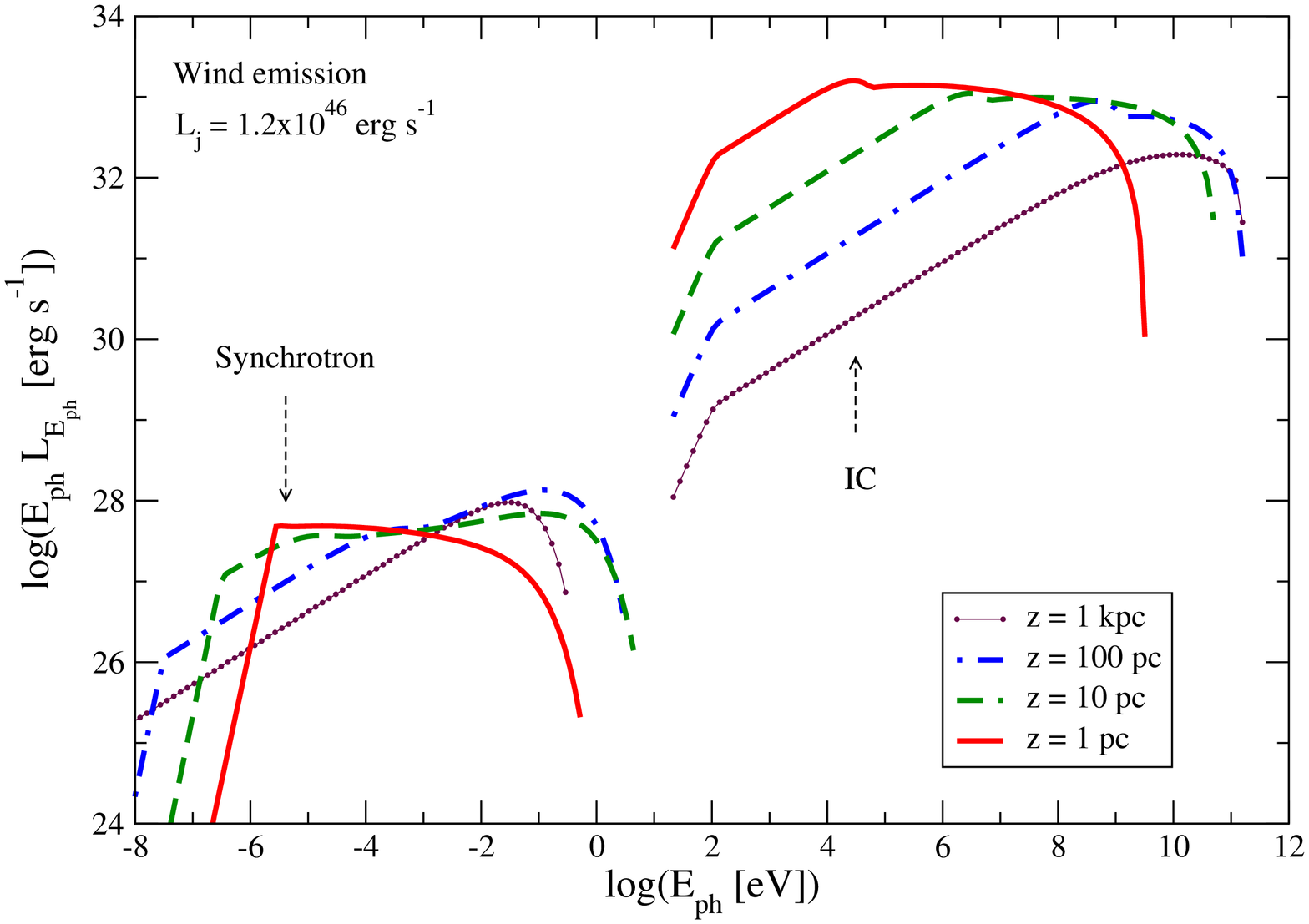}
\caption{Spectral energy distribution produced in the jet
(left) and in the wind (right) by the interaction of only 
one star with a jet of $L_{\rm j} = 1.2\times10^{46}$~erg~s$^{-1}$ at different
$z$.}
\label{sed}       
\end{figure*}

As is shown in Fig.~\ref{sed},
the emission produced by massive stars per interaction at small values of 
$z$ is higher
than emission produced at larger $z$, as a consequence of the dilution
of the target fields (the photon density decreases as 
$z^{-2}$ and $B_{\rm jbs}$ as $z^{-1}$). 
Synchrotron emission produced in the jet bow-shock is more than 100 times 
larger than the emission produced in the wind, but IC radiation generated 
in the jet and in the wind both reach the similar luminosity along $z$. 

\begin{figure}
\centering
\includegraphics[angle=270, width=0.5\textwidth]{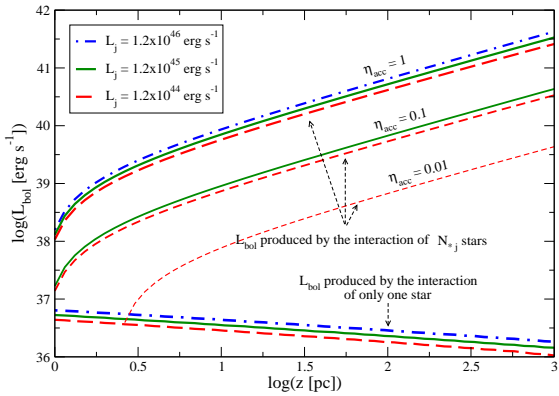}
\caption{Total bolometric luminosities at different $z$.}
\label{Lbol}       
\end{figure}

However, at large values of $z$ the number of stars
interacting with the jet is $> 1$ and the  emission produced by all of them
increases with $z$.
We calculate the emission produced by each one of the massive 
stars located inside the jet at each $z$ and then we integrate along $z$ 
all the contributions, as is shown in Fig.~\ref{Lbol}. The bolometric 
luminosities plotted in the figure correspond to the total emission produced
by each jet/star interactions, i.e. it is the sum of the bolometric
luminosity produced both in the jet and in the wind, by synchrotron and
IC emission. Note that the bolometric luminosities produced by
only one star interacting with the jet at each $z$ are small. However,
when we consider that $N_{\star \rm j}$ stars are inside the jet, the 
bolometric luminosities produced by all of them is significantly larger.

\subsection{Gamma-ray emission}

From Fig.~\ref{sed} we can see that most of the energy radiated by jet/star
interactions at any $z$ is in the gamma domain, in particular in the case
of the wind emission. The jet emission at $z \gtrsim 10$~pc
is dominated  by synchrotron radiation, but at $z \lesssim 10$~pc
synchrotron and IC emission levels are similar.  
In the cases with low values of $N_{\rm j \star}$ ($\eta_{\rm acc} = 0.01$), 
the produced high-energy emission 
can not be detected by  \emph{Fermi} satellite (in the range 0.1-1~GeV).
The most interesting cases are
those with $\eta_{\rm acc} = 0.1$ and $1$, whose emission could be marginally 
detectable in the case of sources located  at
a distance $\lesssim 50$~Mpc. We note that the emission produced by jet/star
interactions  will be more
prominent in AGNs with dense stellar populations. In particular, the 
interaction of a star forming region with a jet will be study by us
in a future work.  

 Given the typical stellar
photon energy $E_0 \sim 10$~eV, gamma rays beyond $\sim 30$~GeV can be
affected by photon-photon absorption due to the presence of
the stellar radiation field.  However, this process is only
important at $z < 1$~pc, where $R_{\rm sp}$ is  small.   
Another effect that
should be considered at energies beyond 100~GeV is absorption
in the extragalactic background light via pair creation (important only
for sources located well beyond 100~Mpc).

\section{Discussion}

We have studied the interaction of massive stars with
AGN jets, focusing on the production of gamma rays
from particles accelerated in the double bow-shock structure formed
around the stars as a consequence of the jet/stellar wind
interaction. We calculated the energy distribution of electrons 
accelerated in the jet and in the wind, and the
subsequent non-thermal emission from these relativistic particles. In
the jet and wind shocked regions, the most relevant radiative processes are
synchrotron emission and IC scattering of stellar photons. The properties of
the emission generated in the downstream region of the bow shocks
change with $z$. On the one hand, the target densities for radiative
interactions decrease as $z^{-2}$.  On the other hand, the time of the
non-thermal particles inside the emitter is $\propto R_{\rm sp}\propto z$, 
and the number of stars per jet length unit 
${\rm d}N_{\star \rm,j}/{\rm d}z\propto z$.
Therefore, for a population of stars, the last two 
effects soften the emission drop with $z$.
In the case of $M_{\rm bh} = 10^{9}$~M$_{\odot}$,
and high accretion rates ($\eta_{\rm acc} = 1$),
gamma-ray luminosities $\sim  10^{41}$~erg~s$^{-1}$ may be
achieved (see Fig.~\ref{Lbol}). However, note that few 
powerful Wolf-Rayet stars inside the jet 
could actually dominate over the whole main-sequence OB star population
\citep{Araudo_12, Araudo_13}. 

Since jet-star emission should be rather isotropic, it would be masked
by jet beamed emission in blazar sources. Although radio loud AGN
jets do not display significant beaming, these
objects may emit gamma-rays from jet/star interactions. 
Misaligned AGNs represent an increasing population of GeV sources, as
is shown is the second catalog of the \emph{Fermi} satellite.
Close and powerful
sources could be detectable by deep enough observations of
\emph{Fermi}. After few-year exposure times, a significant 
signal from these
objects could be found, and their detection can shed light 
on the jet matter composition
as well as on the stellar populations in the vicinity of AGNs.

\section*{Acknowledgments}
This project is financially supported by PAPIIT, UNAM;
PIP 0078/2010 from CONICET,
and PICT 848/2007 (Argentina).
G.E.R. and V.B-R. acknowledge support from  grants AYA 2010-21782-C03-01 and 
FPA2010-22056-C06-02.

\end{document}